\documentclass[12pt,preprint]{aastex}

\usepackage{graphicx}
\usepackage{pdflscape}

\slugcomment{The Astrophysical Journal}

\shorttitle{Episodic Ejection}
\shortauthors{Jewitt et al.}

\begin{document}

\title{Episodic Ejection from Active Asteroid 311P/PANSTARRS 
}
\author{David Jewitt$^{1,2}$, Jessica Agarwal$^3$, Harold Weaver$^4$,  Max Mutchler$^5$ and Stephen Larson$^6$
}
\affil{$^1$Department of Earth and Space Sciences,
University of California at Los Angeles, \\
595 Charles Young Drive East, 
Los Angeles, CA 90095-1567\\
$^2$Department of Physics and Astronomy,
University of California at Los Angeles, \\
430 Portola Plaza, Box 951547,
Los Angeles, CA 90095-1547\\
$^3$ Max Planck Institute for Solar System Research, Justus-von-Liebig-Weg 3, 37077 Gottingen, Germany \\
$^4$ The Johns Hopkins University Applied Physics Laboratory, 11100 Johns Hopkins Road, Laurel, Maryland 20723  \\
$^5$ Space Telescope Science Institute, 3700 San Martin Drive, Baltimore, MD 21218 \\
$^6$ Lunar and Planetary Laboratory, University of Arizona, 1629 E. University Blvd.
Tucson AZ 85721-0092 \\
}

\email{jewitt@ucla.edu}



\begin{abstract}
We examine the development of the active asteroid  311P/PANSTARRS (formerly, 2013 P5) in the period from 2013 September to 2014 February using high resolution images from the Hubble Space Telescope.  This multi-tailed object is characterized by a single, reddish nucleus of absolute magnitude $H \ge$ 18.98$\pm$0.10, corresponding to an equal-area sphere of radius $\le$200$\pm$20 m (for assumed  geometric albedo 0.29$\pm$0.09). We set an upper  limit to the radii of possible companion nuclei at $\sim$10 m.  The nucleus ejected debris in nine discrete episodes, spread irregularly over a nine month interval, each time forming a distinct tail.  Particles in the tails range from about 10 $\mu$m to at least 80 mm in radius, and were ejected at speeds $<$1 m s$^{-1}$.   The ratio of the total ejected dust mass to the nucleus mass is $\sim$3$\times$10$^{-5}$, corresponding to a global surface layer $\sim$2 mm thick, or to a deeper layer covering a smaller fraction of the surface.  The observations are incompatible with an origin of the activity by impact or by the sublimation of entrapped ice.   This object appears to be shedding its regolith by rotational (presumably YORP-driven) instability.  Long-term fading of the photometry (months) is attributed to gradual dissipation of near-nucleus dust.  Photometric variations on short timescales ($<$0.7 hr) are probably caused by fast rotation of the nucleus.  However, because of limited time coverage and dilution of the nucleus signal by near-nucleus dust, we have not been able to determine the rotation period.  
\end{abstract}


\section{Introduction}
Inner main-belt object  311P/PANSTARRS (2013 P5) (hereafter ``311P'')  showed a non-asteroidal appearance at discovery  (Bolin et al.~2013). High resolution images from the Hubble Space Telescope (HST) in 2013 September soon revealed an extraordinary set of six linear, comet-like tails  (Jewitt et al.~2013, hereafter Paper 1).  The orbital semimajor axis, eccentricity and inclination of 311P are 2.189 AU, 0.115 and 5.0\degr, respectively, all typical of the inner, main-belt asteroids. Unlike comets, which have Tisserand parameters relative to Jupiter $T_J <$ 3 (e.g.~Kresak 1982), the parameter for 311P is an asteroid-like $T_J$ = 3.66.  The combination of asteroid-like orbit and comet-like mass-loss together reveal 311P as an active asteroid (Jewitt 2012) or, equivalently, a main-belt comet (Hsieh and Jewitt 2006).  

A  clue concerning the origin of the mass loss in 311P was provided by the tail  position angles, which varied with time in such a way as to show that each tail is a synchrone (Paper 1).  In a synchrone, dust particles having a wide range of sizes are released from the nucleus simultaneously and are then sorted in size by solar radiation pressure (Finson and Probstein 1968).  Synchrone trajectories fitted to the Hubble data showed that tails were formed intermittently from 2013 April to September, with long periods of quiescence in between.  We compared several activity mechanisms (drawn from Jewitt 2012) with the available data.  The episodic ejection of dust over a long time period is inconsistent with a collisional origin. An impact would produce dust impulsively, in a single burst, not in multiple bursts spread over many months.   Ice-sublimating comets are commonly variable in their output, but no comets have shown activity like that of 311P, with brief periods of ejection interspersed with long but irregular intervals of quiescence.  Indeed, the inner-belt location and associated high equilibrium temperatures (the spherical blackbody temperature at 2 AU is 197 K, while the sub-solar temperature is $\sim$279 K) argue against the survival of surface water ice.  In addition, most asteroids with orbits in the vicinity of 311P consist of highly metamorphosed rocks, with meteorite counterparts in the LL chondrites, unlikely to preserve ice (Paper 1).   Other mechanisms, including thermal fracture and electrostatic ejection also offer unlikely sources of mass loss from 311P.  The thermal environment on 311P is unremarkable and benign, relative to the one instance ((3200) Phaethon, whose surface reaches 1000 K when near perihelion) in which this process has been shown to be effective.  Electrostatic forces alone are too weak to eject the large particles present in the debris of 311P.
   
In Paper 1 we concluded  that rotational instability offers a plausible explanation for the activity in 311P, given that the nucleus is sub-kilometer in size and the estimated timescale for spin-up by radiation forces is correspondingly short ($<$1 Myr).  By coincidence, an example of a structural rotational instability, in which the body of an asteroid is observed to be breaking apart, has recently been identified in the multiple object P/2013 R3 (Jewitt et al.~2014).  311P appears qualitatively different in having a single nucleus that emits debris episodically.  It may be an example of a surface or shedding instability, in which loose surficial material is lost owing to rotational instability while the main body remains intact (Hirabayashi and Scheeres 2014).   Our  conclusions about 311P in Paper 1  were drawn from HST observations taken at two epochs, UT 2013 Sep 10 and 23, soon after discovery.  Here, we expand and extend our conclusions using  these  and  new HST observations from five additional dates between 2013 October and 2014 February.

 

\section{Observations} 
\begin{figure}
\epsscale{1.0}
\begin{center}
\includegraphics[width=1.0\textwidth]{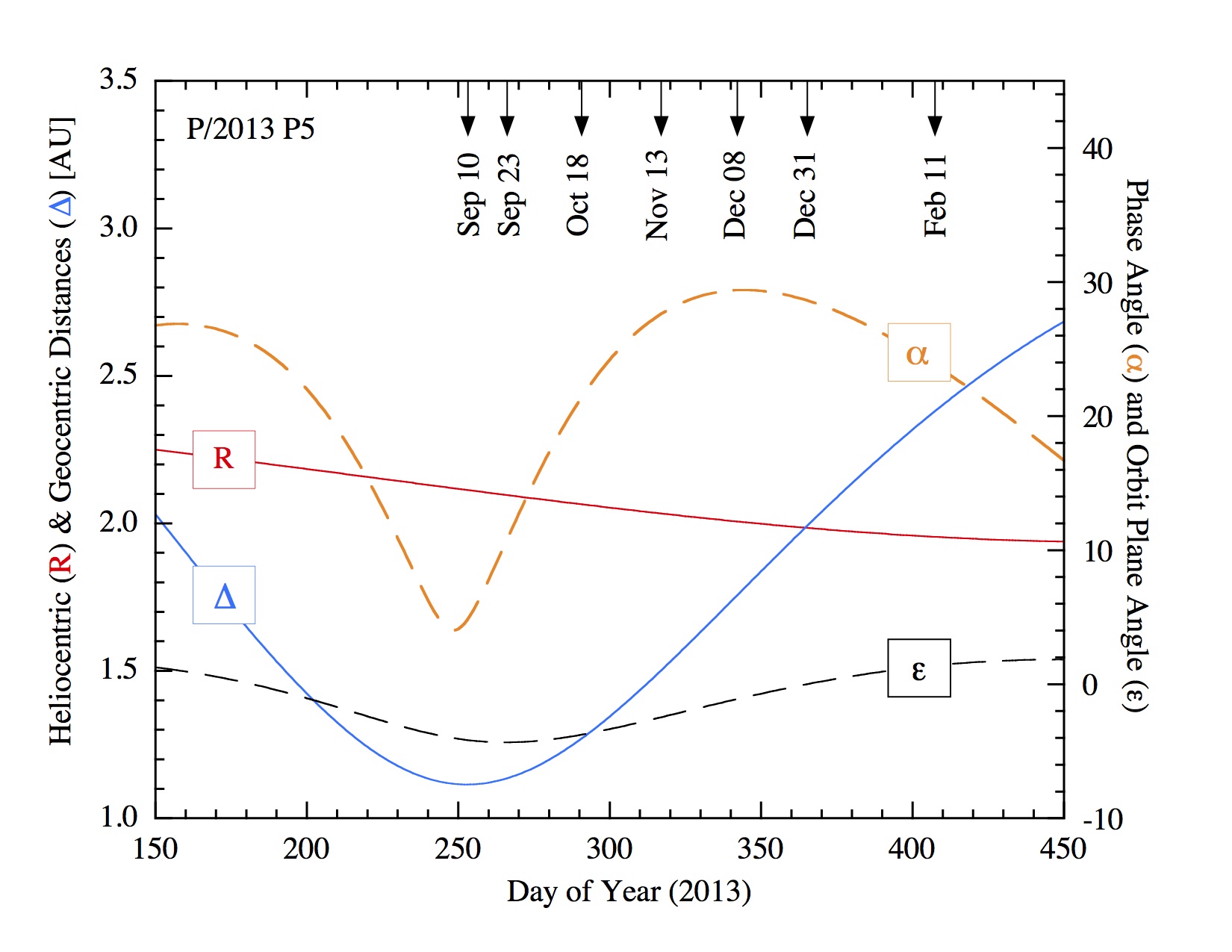}
\caption{Geometric circumstances of the observations (c.f.~Table \ref{geometry}) $R, \Delta, \alpha$ and $\epsilon$ are, respectively, (left axis): the heliocentric and geocentric distances, (right axis): the phase angle and the angle of the Earth above the orbital plane of P/2013 311P. The dates of the HST observations are marked along the upper axis. \label{rDa}
} 
\end{center} 
\end{figure}

New observations from seven orbits of Director's Discretionary Time (General Observer program number 13609) are combined with two orbits obtained under a Target of Opportunity program (General Observer program number 13475). We used the WFC3 camera \citep{2012wfci.book.....D} with 0.04\arcsec~pixels each corresponding to about 33 km at the first observation in September, increasing to about 69 km by the last observation in 2014 February.  The Nyquist-sampled (two-pixel) spatial resolution ranged from 66 to 138 km.  All observations were taken using the extremely broad F350LP filter (full-width-at-half-maximum 4758\AA) which provides maximum sensitivity to faint sources at the expense of introducing some uncertainty in the transformation to standard astronomical filter sets.  The effective wavelength for a solar-type (G2V) source is 6230\AA.  A journal of observations is given in Table \ref{geometry} while Figure (\ref{rDa}) provides a graphical representation.

\begin{figure}
\epsscale{1.0}
\begin{center}
\plotone{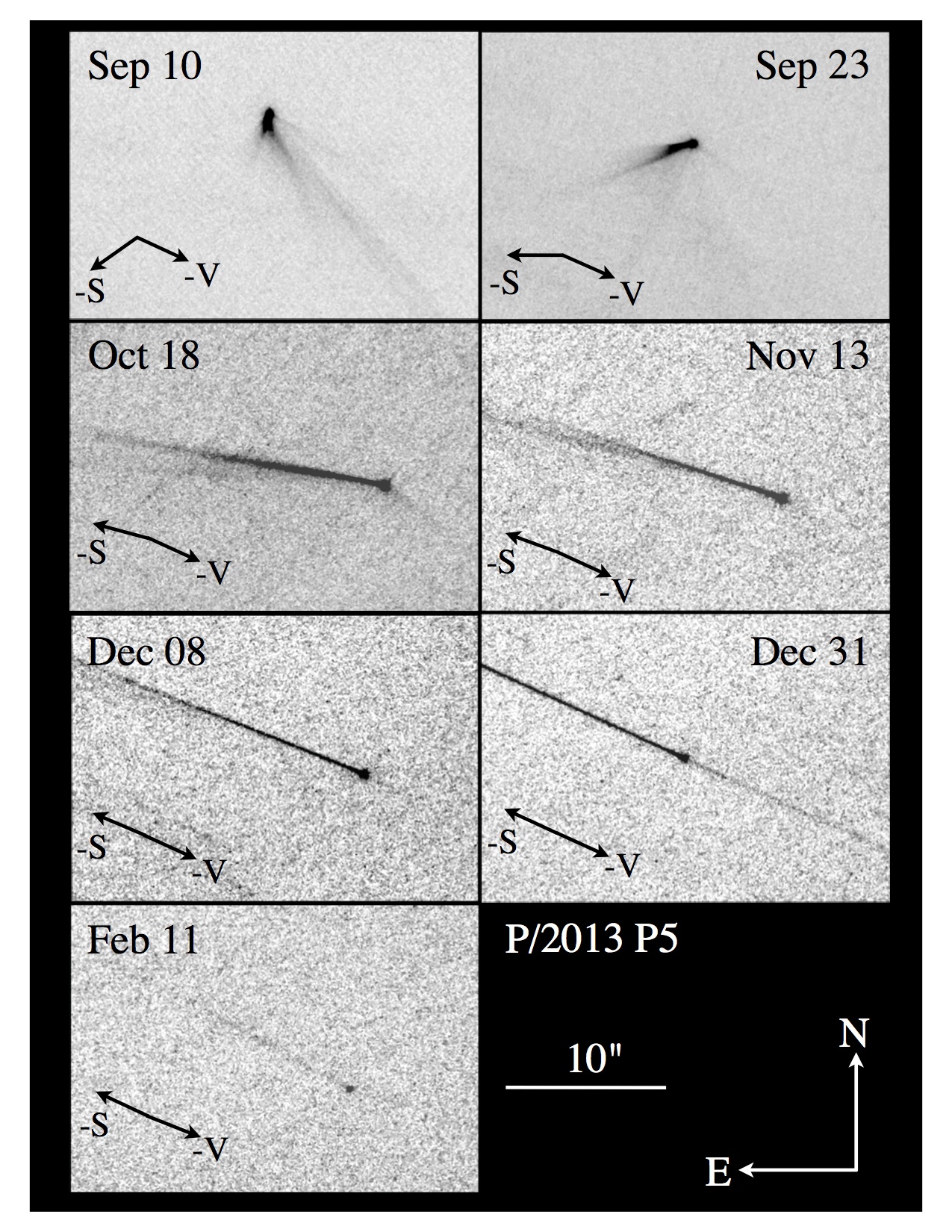}
\caption{Composite of images from the seven epochs of observation, with dates in 2013-2014 marked for each panel.  Arrows show the directions of the negative heliocentric velocity vector (-V) and the antisolar direction (-S). \label{image}
} 
\end{center} 
\end{figure}

In each orbit we obtained five exposures of 348 s and one of 233 s. The drizzle-combined images for each date are shown in Figure (\ref{image}), where they are presented at a fixed angular scale and orientation.  In the Figure are marked the position angles of the antisolar direction and the negative heliocentric velocity vector  (c.f.~Table \ref{geometry}).  Individual and combined images from each orbit were examined to identify the dust tails.  The main difficulty in this exercise was presented by the extreme faintness of many of the tails and, in some cases, by confusion with the trails made by background stars and galaxies.   Field objects trailing parallel to dust tails presented a particular problem which, however, we were able to resolve by comparing different images from within a single orbit.   We  briefly comment on each panel of Figure (\ref{image}).

\textbf{September 10:}  A bright nucleus is the source of six straight tails with position angles splayed over the 141$\pm$2\degr~to 237$\pm$1\degr~range.  The tails are aligned neither with the projected orbit nor with the antisolar direction.

\textbf{September 23:}  The same six tails can be seen again, although more faintly, and with position angles that have rotated clockwise in the plane of the sky.  This rotation provided the main clue that the tails are synchrones (Paper 1), caused by impulsive dust emission on a range of dates.

\textbf{October 18:} Most tails have collapsed into a narrow range of position angles near 80 to 86\degr, except that a short tail near 234\degr~can be seen. 

\textbf{November 13:} The tails to the east of the nucleus continue to compress into a narrower range of position angles, while the sunward tail is extremely faint.

\textbf{December 08:}  Similar to November 13.

\textbf{December 31:} This observation was specially scheduled to coincide with the passage of the Earth through the orbital plane of 311P.  From this vantage point, the apparent extent of the object in the direction perpendicular to the projected orbit is a measure of the true vertical extent, free from the confusing effects of projection.  We used this geometry to measure the normal component of the dust ejection velocity.  The composite image shows dust along the projected orbit to both the east and west sides of the nucleus.

\textbf{February 11:} Now very faint because of the increased geocentric distance (Table \ref{geometry}), the nucleus is seen attended by a single, low surface brightness tail extending in the antisolar direction.


\section{Nucleus}

The principal impediments to accurate photometry with WFC3  data are background objects (stars and galaxies) smeared across the images by the parallactic motion of the telescope and a high flux of charged particles (``cosmic rays'') impinging on the CCD.  To the extent that field objects and cosmic rays are spatially uncorrelated, most can be readily removed by carefully combining the six images taken within a single orbit.  However, for photometry using individual images the contaminating objects must be removed by another method.  We used spatial interpolation across affected pixels to remove cosmic rays.  Special attention was paid to the near-nucleus region, where cosmic rays are hard to identify because of the steep surface brightness gradients.  To identify cosmic rays in this region, we first computed the drizzle-combination of the images from a whole orbit and then subtracted this from each individual image, leaving only noise and cosmic rays.  After removing the latter by hand, the drizzle-combination image was then added back to each individual image.  We used simple aperture photometry on the resulting corrected images to estimate the brightness of the nucleus region.  

\subsection{Nucleus Size}
We employed a circular aperture of projected angular radius 0.2\arcsec~with background subtraction from a concentric annulus having inner and outer radii 1.0\arcsec~and 2.0\arcsec, respectively.  The  aperture admits a small contribution from near-nucleus dust; our attempts to remove this using modeling of the point-spread function of the data proved unreliable.  We used the HST exposure time calculator to convert the measured count rate into an effective V magnitude, finding that a V = 0 G2V source gives a count rate of 4.72 $\times$10$^{10}$ s$^{-1}$ within a 0.2\arcsec~radius photometry aperture.   The photometric results are summarized in Table (\ref{photometry}).  In the five months between UT 2013 September 10 and 2014 February 11, the nucleus magnitude faded from $V_{0.2}$ = 20.92$\pm$0.01 to 23.30$\pm$0.02.  A large fraction of this fading can be attributed to the changing geometry of observation.

\begin{figure}
\epsscale{1.0}
\begin{center}
\plotone{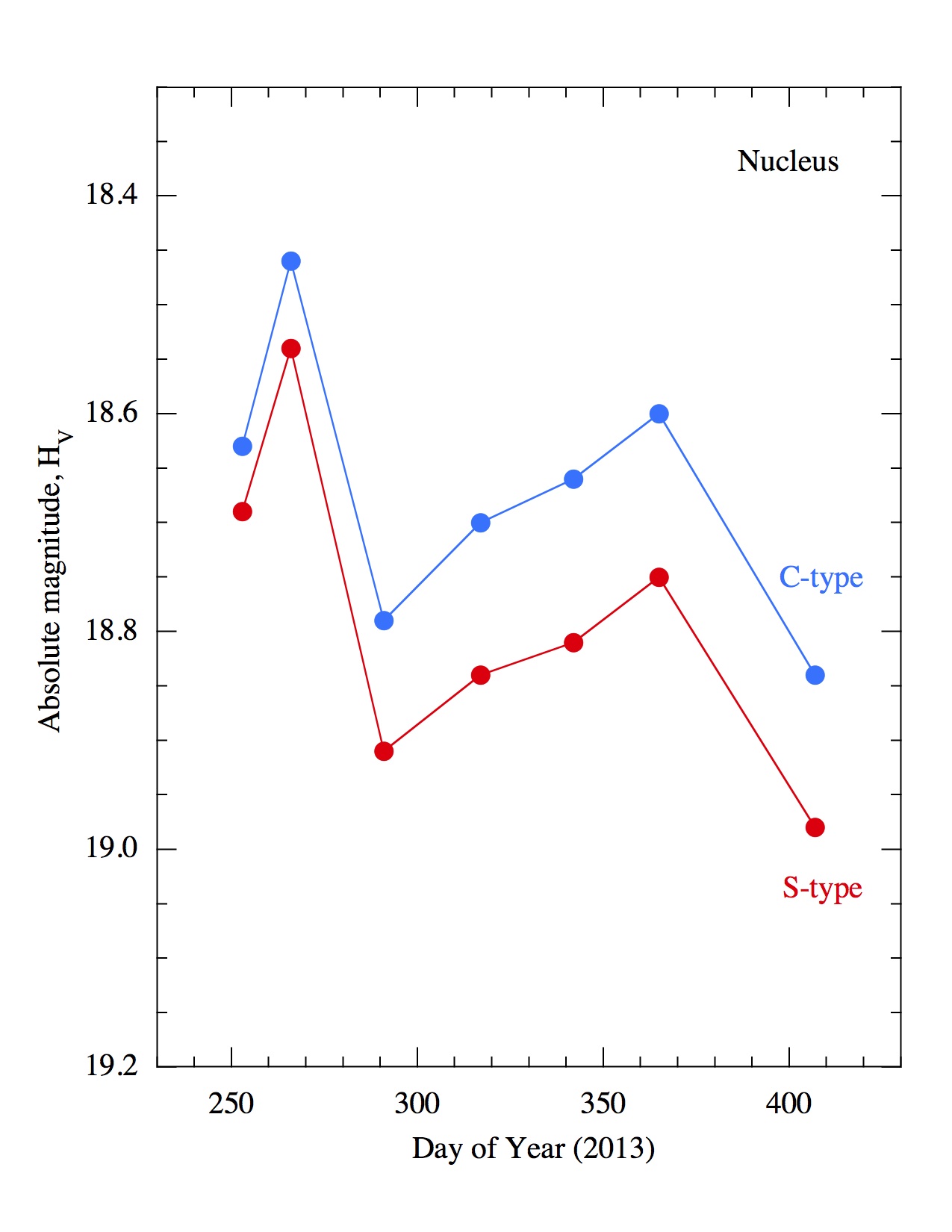}
\caption{Absolute magnitude of the nucleus vs.~day of year (Table \ref{photometry}).  Red and blue circles, respectively, show the median magnitude at each date, corrected using the phase functions of S-type and C-type asteroids.   Error bars on the photometry are about the same size as the plotted symbols.    \label{H_vs_DOY}
} 
\end{center} 
\end{figure}

The absolute magnitude (i.e.~corrected to unit heliocentric and geocentric distances, and to zero phase angle) was computed from

\begin{equation}
H_V = V - 2.5 \log_{10}\left(R^2 \Delta^2\right) +  2.5 \log_{10}(\Phi(\alpha))
\label{H_v}
\end{equation}

\noindent where the phase function, $\Phi(\alpha)$, is the ratio of the brightness of the object at phase angle $\alpha$  to the brightness at $\alpha$ = 0\degr.  The phase function is unmeasured for 311P, but has been assessed for a large number of asteroids of different spectral types.  We use the so-called HG phase functions for C-type ($g$ = 0.15) and S-type ($g$ = 0.25) asteroids from Bowell et al.~(1989) to bracket the phase dimming in 311P.  Values of $H_V$ computed using these functions are shown in Figure (\ref{H_vs_DOY}) vs.~the date of observation.  Data points in the figure represent the median magnitude on each date, with statistical uncertainties ($\sim \pm$0.01 mag.) that are comparable in size to the symbols in the plot.  The uncertainties due to the unknown phase function are larger than the  photometric uncertainties, reaching about 0.15 magnitudes at the largest phase angles. The color of 311P suggests that it is closer to S-type than C-type, as does its orbital location in the inner belt, where S-types are common (Paper 1).  If so, the absolute magnitude varies from $H_{0.2}(S)$ = 18.54$\pm$0.01 on UT 2013 Sep 23 to  $H_{0.2}(S)$ = 18.98$\pm$0.02 on UT 2014 Feb 11, a range of $\sim$0.5 magnitude.    This range reflects changes in the dust content of the 0.2\arcsec~photometry aperture, with later measurements generally showing less evidence for near-nucleus dust contamination.  While we cannot be sure that the bare nucleus has been measured, even in the February 11 data, we can safely take  $H_{V}$ = 18.98$\pm$0.10 (where the enlarged uncertainty is an attempt to represent the uncertainty introduced by the unmeasured phase function) as our upper limit to the possible brightness of the nucleus.  

The  physical properties of 311P are related to $H_{V}$ by

\begin{equation}
p_V C_e = 2.24\times 10^{22} \pi 10^{0.4(V_{\odot} - H_V)}
\label{invsq}
\end{equation}

\noindent in which $V_{\odot}$ = -26.75 is the V magnitude of the Sun (Drilling and Landolt 2000), $p_V$ is the geometric albedo and $C_e$ is the geometric cross-section.  As noted in Paper 1, the orbit of 311P is close to the Flora asteroid family, for which the mean geometric albedo is $p_V$ =  0.29$\pm$0.09 (Masiero et al.~2013).   Substitution of this albedo into Equation (\ref{invsq})  gives $C_e$ = 0.12$\pm$0.01 km$^2$, corresponding to a spherical nucleus of effective radius $r_e = (C_e/\pi)^{1/2}$ = 0.20$\pm$0.02 km. A smaller albedo would increase $r_e$ relative to this value while the possibility of near-nucleus dust contamination of the photometry aperture would decrease it.  The mass of a 200 meter radius sphere of nominal density $\rho$ = 3300 kg m$^{-3}$ (the  average density of LL chondrites according to Consolmagno et al.~2008) is $M_n$ = 1.1$\times$10$^{11}$ kg and the gravitational escape speed $v$ = 0.27 m s$^{-1}$.

\subsection{Nucleus Rotation}
\begin{figure}
\epsscale{1.0}
\begin{center}
\plotone{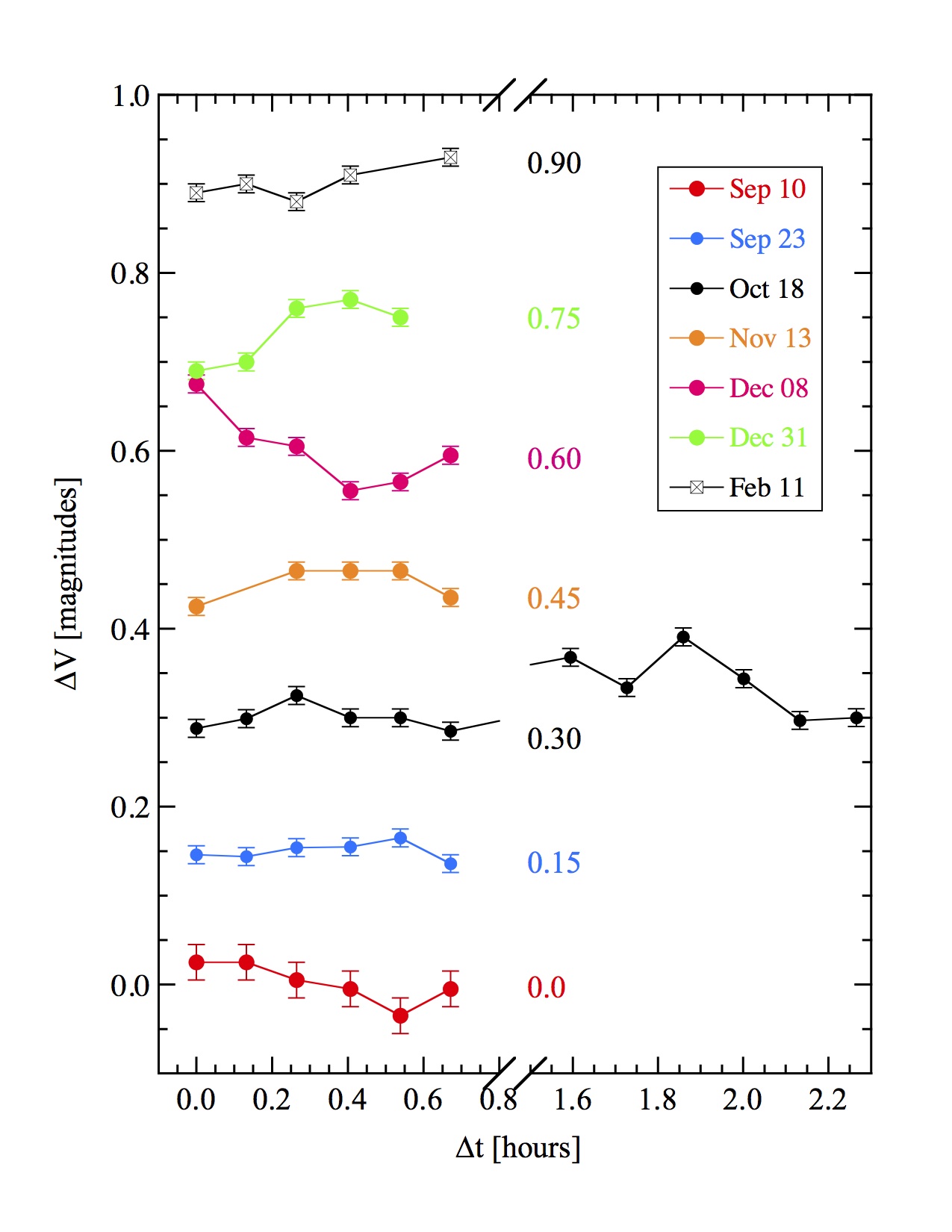}
\caption{Short-term brightness variations from each visit.  The plotted magnitudes are $\Delta V = V - V_m + C$, where $V_m$ is the median magnitude within each visit and C is an arbitrary constant, as indicated in the plot.  \label{DV_Dt}} 
\end{center} 
\end{figure}

We examined the six images from each orbit individually to search for temporal variation that might result from rotation of an irregular nucleus.  Figure (\ref{DV_Dt}) shows the resulting short-term photometric variability of the nucleus of 311P.   For ease of plotting the data, we have scaled the time using $\Delta t_i = t_i - t_1$,  where $t_i$ is the start time of the $i$th image taken in each orbit.  Additionally, we computed the reduced magnitude $\Delta V = V_i -V_m +C$, where $V_i$ is the apparent magnitude in the $i$th image, $V_m$ is the median apparent magnitude within a given orbit and $C$ is a constant.  The latter was picked simply to space the lightcurves for clarity in Figure (\ref{DV_Dt}).  

Evidently, the nucleus region of 311P does vary in brightness even within the $\sim$0.7 hour duration of a single HST orbit.  The change in magnitude is $\sim$0.1 on 2013 October 18, December 08 and 31, and smaller on the other dates of observation, but still larger than the $\pm$0.01 magnitude statistical uncertainties.  It is unlikely that the observed short-term variations are due to changes in the rate of production of near-nucleus dust, because the aperture crossing times are too long.  To see this, note that at geocentric distances $1.1 \le \Delta \le 2.4$ AU (Table \ref{geometry}), the linear radius of the 0.2\arcsec~aperture projected to 311P is $160 \le \ell \le 350$ km.  With nominal  dust ejection velocity $v_d$ = 0.5 m s$^{-1}$ (determined from the dust dynamics in Section 4.3), we calculate an aperture crossing time $t = \ell / V_d$ in the range $3.7 \le t\le 8.1$ days.  This is $\sim$10$^2$ times longer than the timescale of the observed brightness variations.  We conclude that the short-term brightness fluctuations are not caused by dust.  On the other hand,  variable amounts of near-nucleus dust provide a natural explanation for observed variations in the mean brightness  by up to $\sim$0.5 magnitudes on longer ($\sim$month) timescales  (Table \ref{photometry} and Figure \ref{H_vs_DOY}).  

We interpret the short-term ($\sim$hour) photometric variations as the result of nucleus rotation, while the longer-term ($\sim$month) variations as produced by the variable dust production rate.  We sought but did not find the rotation period of the nucleus.  The failure is easily understood because 1) the photometry is heavily undersampled (HST visits lasting $\sim$0.7 hr are spaced at intervals of months; c.f.~Table \ref{geometry}) and 2) variable quantities of near-nucleus dust presumably alter the amplitude of photometric variation from month-to-month by dilution.  In addition, 311P could be in an excited (non-principal axis) rotational state, leading to a lightcurve that is not singly periodic and which is therefore more difficult to determine from limited data.   The data suggest that the apparent rotational range of the lightcurve is $\lesssim$0.15 magnitudes but, again,  it is difficult to relate this range to the nucleus axis ratio because of possible dilution by light scattered from near-nucleus dust.   Nevertheless, the existence of rapid photometric variations is consistent with the rapid rotation of the nucleus.  

Hainaut et al.~(2014) have suggested an alternate scenario in which 311P is a newly-forming contact binary consisting of two prolate shaped components rubbing together at their tips.  Grinding at the contact points of the binary releases dust.  Their scenario predicts that the rotational range of the lightcurve should be large and that, for  values of the nucleus density in the range 1000 to 3000 kg m$^{-3}$, the rotation period should be 6.7 to 11.6 hr.  Neither the small measured photometric range nor the observed rapid lightcurve variations  (Figure \ref{DV_Dt}) provide support for this possibility.

\subsection{Limits to Companions}
Motivated by the multiple nucleus appearance of P/2013 R3 (Jewitt et al.~2014), we searched for companion nuclei in the vicinity of 311P.  For this purpose, we took scaled versions of the 311P composite images (total exposure 1973 s), offset them and added them back to the original image.  The morphological complexity and temporal variability of 311P make it impossible to specify with accuracy a single limit to the brightness of possible companion objects.  Early observations suffer from a large quantity of ejected dust, which creates a high surface brightness background in the near-nucleus region and limits sensitivity to any faint companions (c.f.~Figure \ref{image}).  Later observations suffer less from dust but the sensitivity to faint companions is diminished by the inverse square law and the increasing distance to the object (Table \ref{geometry}).  As a compromise, we used data from UT 2013 October 18 to find that point sources with $V <$ 28.5 are readily apparent if projected against the sky background $\ge$0.4\arcsec~(370 km) from the nucleus of 311P.  This empirically determined  limit matches the expected sensitivity of WFC3 to point sources as determined from the on-line Space Telescope Exposure Time Calculator (for $V$ = 28.5, the latter gives an expected signal-to-noise ratio of 2.5 when applied to a spectral class G2V point source).  By Equations (\ref{H_v}) and (\ref{invsq}), $V$ = 28.5 corresponds to a spherical body of radius 10 m, assuming the same S-type phase function and albedo as for the main nucleus.  A larger object could go undetected if its angular distance from the nucleus were $<$0.4\arcsec~or if projected against the brightest dust tails of 311P, but it is unlikely that such a companion could remain hidden in observations over many months, given the changing observing geometry.  We can confidently reject the possibility that 311P is like P/2013 R3, in which there are four large nuclei of $\lesssim$0.2 km scale (Jewitt et al.~2014a).  The mass that would be contained in a 10 m body is 10$^{-4}$ of the mass of the 200 m primary, which sets an absolute upper limit to the ejected mass fraction.  In section \ref{quantity}, we find a smaller value of the ejected mass by examining the  dust.

\section{Dust}
\subsection{Dust Production}
We determined  the position angles  of the 311P dust tails by measuring the brightest pixel in each tail as a function of distance from the nucleus and then fitting a straight line to the measurements by least squares.   The least-squares fits also provided an estimate of the uncertainty on the position angle for each tail.  In some observations,  field stars trailed by parallax  left image residuals that resembled real dust structures in 311P.  We were able to reject these cases by comparing images constructed using different subsets of the data, since the parallax angle changes rapidly as HST moves around its orbit.  The resulting position angle determinations are summarized with their uncertainties in Table (\ref{angles}).

\begin{figure}
\epsscale{1.0}
\begin{center}
\includegraphics[width=0.85\textwidth, angle =90 ]{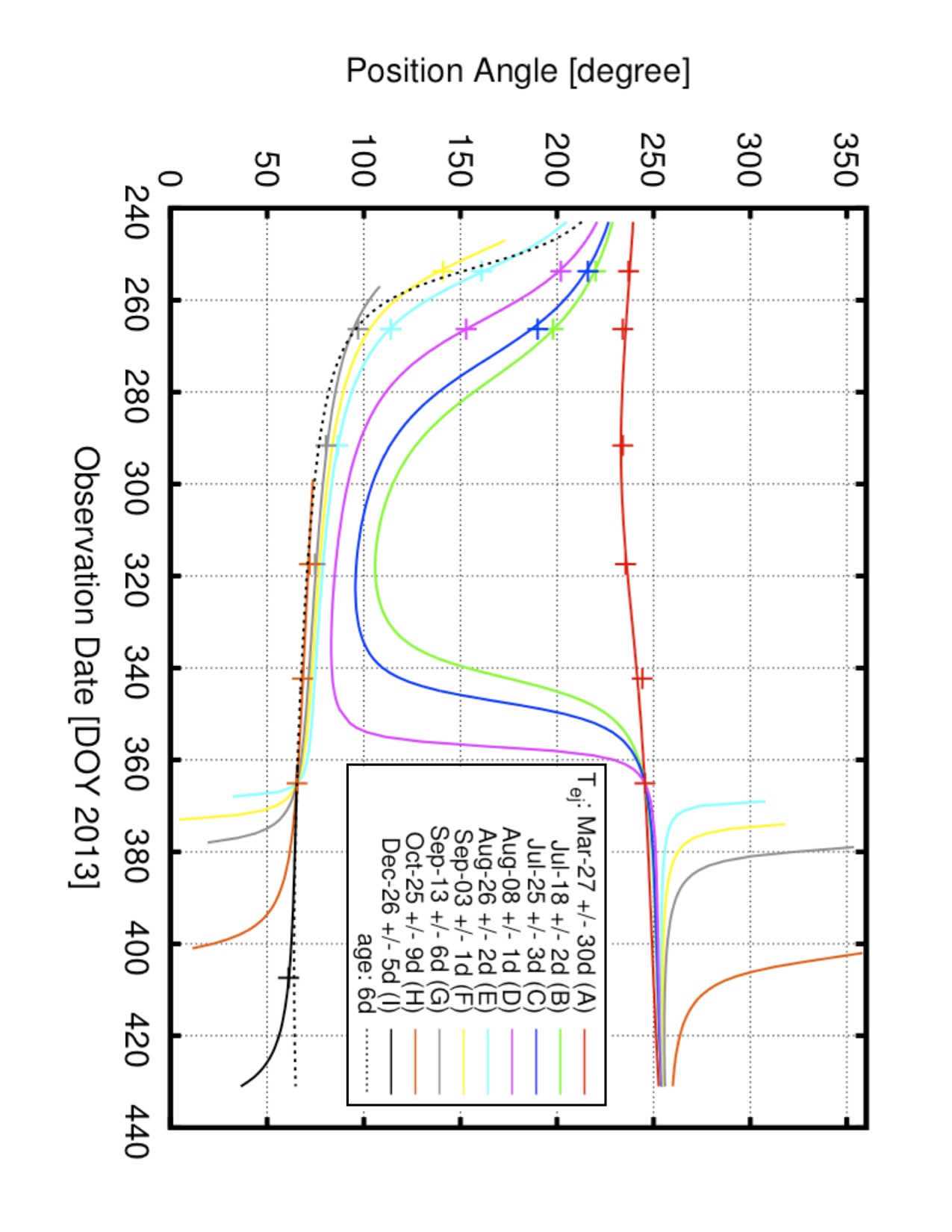}
\caption{ Position angles of the tails from Table \ref{angles} (symbols) and calculated synchrones (solid lines) as functions of the date of observation.  Uncertainties on the measured position angles are  too small to be seen at the scale of the figure, but are listed in Table (\ref{angles}).  The synchrone initiation times, $t_0$, are listed. \label{synchrones}} 
\end{center} 
\end{figure}

\begin{figure}
\epsscale{1.0}
\begin{center}
\plotone{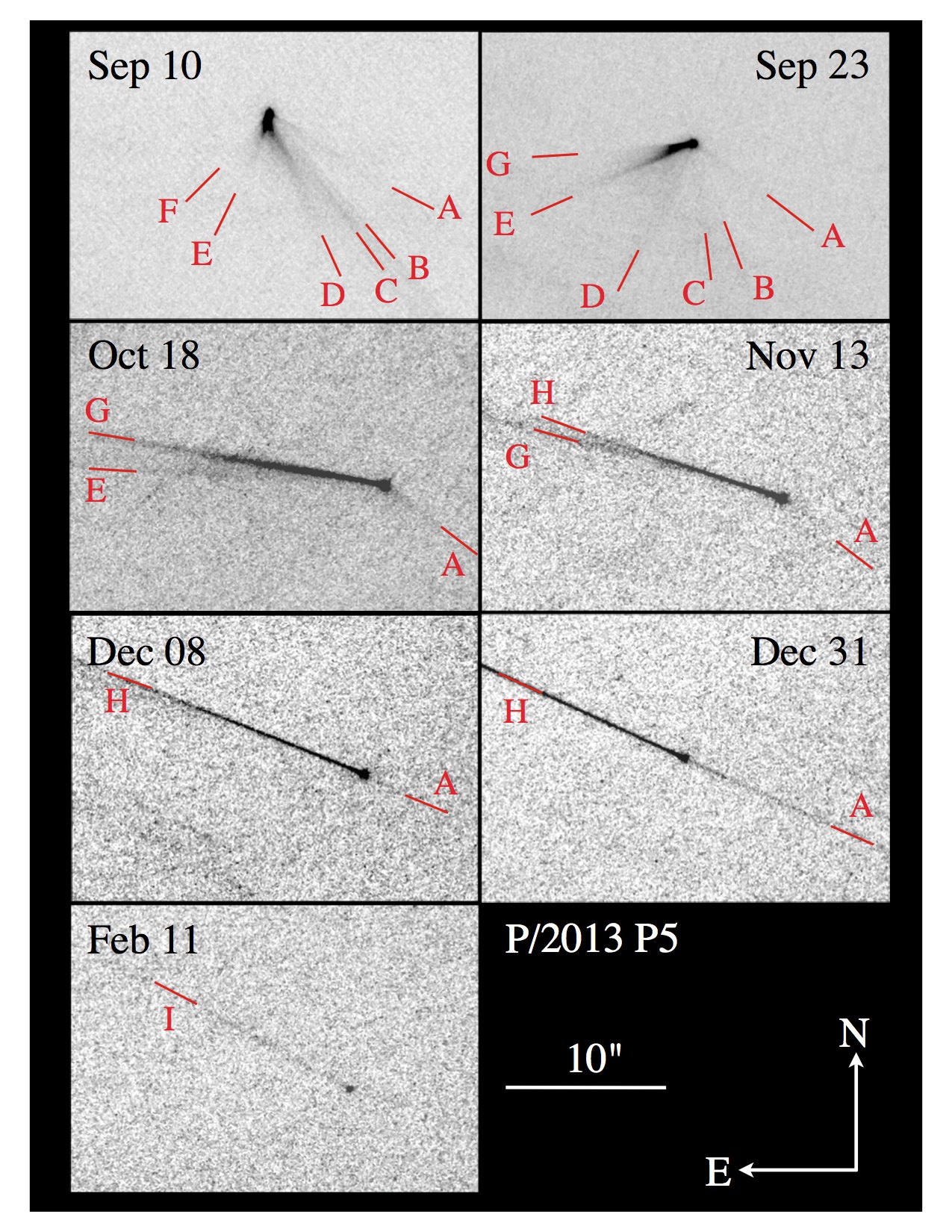}
\caption{ Same as Figure (\ref{image}) but showing identification of the dust tails based on the synchrone fits illustrated in Figure (\ref{synchrones}). \label{image_arrows}
} 
\end{center} 
\end{figure}

A synchrone model (Finson and Probstein 1968) was used to fit the position angle measurements.  In this model, dust grains having a wide range of sizes are assumed to be released from the nucleus with zero initial relative velocity and are then accelerated by solar radiation pressure to form a tail.  The magnitude of the solar radiation pressure acceleration is approximately inversely proportional to the dust grain size.  Therefore, particles in each tail are sorted by radiation pressure such that the smallest are, in a given time, pushed far from the nucleus while the largest ones loiter close to it. The position angle of each tail as a function of time is controlled by a single parameter;  the date of particle release.  Table (\ref{angles}) lists the  synchrone initiation dates derived from the model fits.  Figure (\ref{synchrones}) shows the measured tail position angles and the synchrone fits, with ejections expressed as calendar dates and  observation dates in DOY (Day of Year).  The tails identified from the synchrone fits are marked in Figure (\ref{image_arrows}).

 In general, there is good  agreement between the synchrone solutions deduced from only two observations in Paper 1 and those deduced from the full dataset here.  The most consistently observed tail structure is Tail A, present in all data from 2013 September 10 to 2013 December 31.  The position angles of Tail A indicate ejection on UT 2013 Mar 27 (DOY 86), about five months prior to the discovery epoch (DOY 230) and six months prior to the first HST observation (DOY 253).  Particles in Tail A are comparatively large and slow, lingering close to the projected orbit of 311P.  The initiation date for this tail is the least well-determined, having an estimated uncertainty of $\pm$30 days.  Initiation dates of the other tails are accurate to within a few days, as listed in Figure (\ref{synchrones}).  Tail F in Paper 1 was relatively poorly fitted by a single synchrone (see Figure 2 of Paper 1).  The addition of more data shows that this tail is actually the superposition of two tails, now called F and G, with initiation dates separate by about 10 days.   The least well observed tails are F and I, each observed on only one occasion.  Other tails were detected three times (E, G and H) and two times (B, C and D).  The progressive disappearance of the tails (other than Tail A) on timescales $\sim$1 to 2 months appears to reflect a real depletion of the dust by radiation pressure.    Moreno et al.~(2014) and Hainaut et al.~(2014) confirmed the results of Paper 1 in analyses which supplemented our September 10 and 23 HST visits with lower resolution ground-based data from September and October.

\subsection{Quantity of Dust}
\label{quantity}
The most direct measure of the total quantity of dust in 311P is provided by the integrated photometry obtained using a large projected aperture.  Large aperture photometry  suffers primarily from uncertainties introduced by the non-uniform sky background.  The HST images are susceptible to background contamination by stars and galaxies which are swept over large distances in the image plane by the parallactic motion of the telescope.  For example, imperfectly removed  field objects are obvious as diagonal streaks in the October 18 and November 13 panels of Figure (\ref{image}).  By trial and error, we found that a circular aperture 6.0\arcsec~in radius, with sky subtraction from a surrounding annulus extending to 12.0\arcsec, gives the best measure of the total light from 311P on UT 2013 September 10.   On other dates, we scaled the radius of the photometry aperture in inverse proportion to the geocentric distance (Table \ref{geometry}) in order to ensure that a fixed volume around the nucleus was always measured.  The background determination was made in a contiguous annulus with the difference between the inner and outer radii fixed at 6.0\arcsec.

\begin{figure}
\epsscale{1.0}
\begin{center}
\plotone{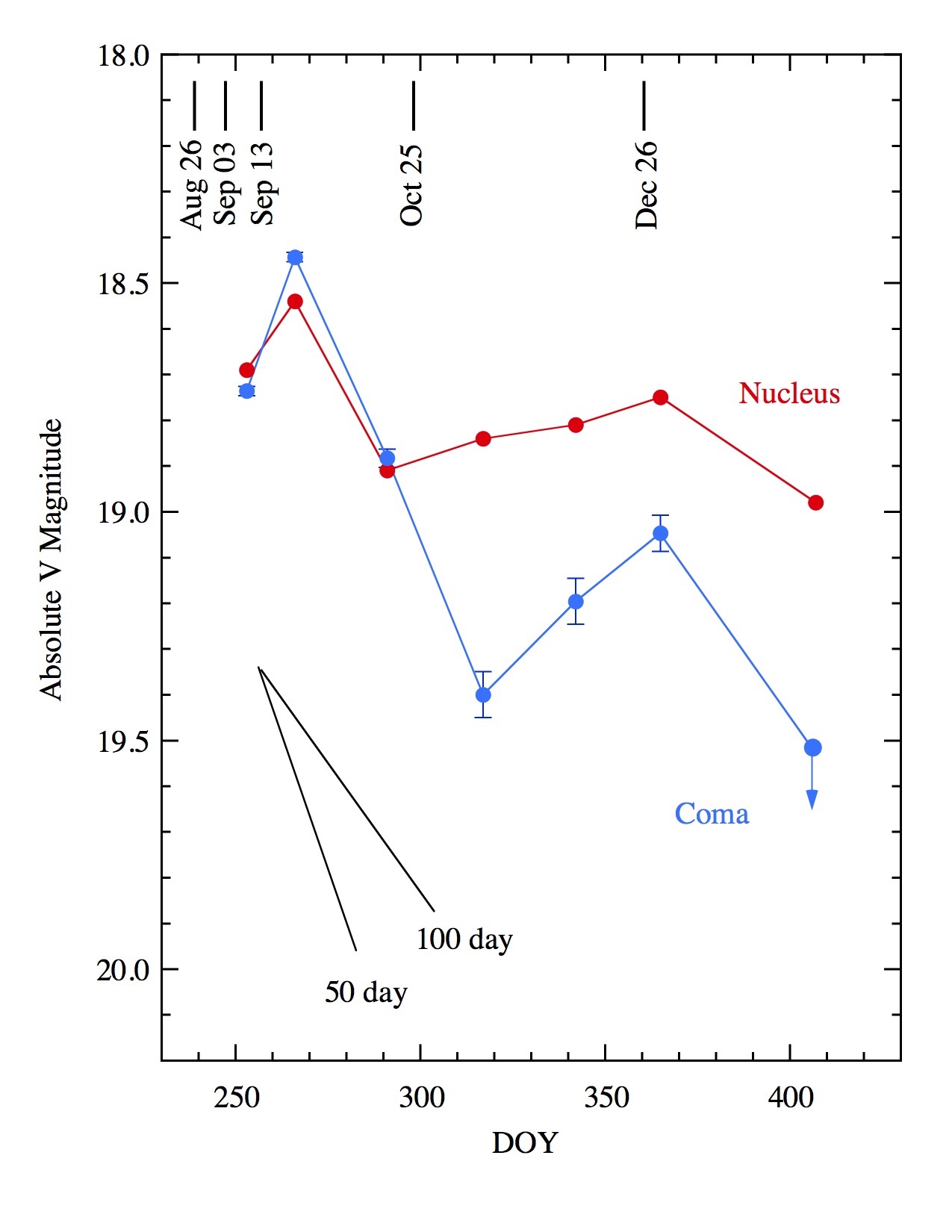}
\caption{Absolute magnitude of the nucleus (red) and coma (blue) vs.~day of year (data from Tables \ref{photometry} and \ref{total_photometry}).  An S-type phase function was assumed for both.  Diagonal lines indicate e-folding decay times of 50 and 100 days, as marked.  The dates of tail initiation events deduced from synchrone analysis of the tail position angles (Table \ref{angles}) are shown.   \label{nucleus_coma}} 
\end{center} 
\end{figure}

These measurements, $V_{\theta}$, are summarized in Table (\ref{total_photometry}) together with our estimates of the errors attributable to background subtraction.  We found that the difference between photometry obtained using a fixed aperture of 6.0\arcsec~radius and one scaled inversely with the geocentric distance was $\le$0.1 magnitudes. Note that Table (\ref{total_photometry}) supersedes large aperture measurements from Table 2 of Paper 1, which are in error because of incorrect background subtraction by the first author.  We also list in Table (\ref{total_photometry}) the absolute magnitude calculated from $V_{\theta}$ using the inverse square law and an assumed S-type asteroid phase function, $H_{\theta}(S)$.   A fraction of the light in the large-aperture photometry is scattered from the nucleus.  To isolate the contribution from the coma, we computed 

\begin{equation}
H_{dust} = -2.5\times log_{10}\left[10^{-0.4 H_{\theta}(S)} - 10^{-0.4 H_{0.2}(S)}\right]
\label{dust}
\end{equation}

\noindent with $H_{0.2}(S)$ from Table (\ref{photometry}) and $H_{\theta}(S)$ from Table (\ref{total_photometry}).  Values of $C_e$ computed from $H_{dust}$ and Equation (\ref{invsq}) using $p_V$ = 0.29 are given in the last column of Table (\ref{total_photometry}).  The cross-sections vary from a peak $C_e$ = 0.2 km$^2$ on September 23 to a minimum $C_e <$ 0.07 km$^2$ on February 11.  However, the dust cross-sections do not fall monotonically as would be expected from dust produced impulsively, for example, by an impact.  Instead, the dust cross-section has a local maximum in the UT 2013 September 23 observations, apparently due to the release of fresh material associated with the production of Tail G.  Between September 23 (DOY 266) and October 18 (DOY 291), Figure (\ref{nucleus_coma}) shows that the coma cross-section steadily decreases, with an e-folding time for the brightness of $\tau \sim$50  days (4.3$\times$10$^6$ s).  If this is a measure of the residence time of the dust in the projected annulus, we can infer an effective dust speed 

\begin{equation}
v_d \sim \frac{7.3\times 10^5 \theta \overline{\Delta} }{\tau}
\end{equation}

\noindent where we take $\theta$ = 5.6\arcsec~as the average of the aperture radii on September 23 and October 18 (Table \ref{total_photometry}) and $\overline{\Delta}$ = 1.21 AU is the average geocentric distance  (Table \ref{geometry}).  Substituting for $\tau$, we find $v_d \sim$ 1 m s$^{-1}$.  This is a strong upper limit to the dust ejection velocity from the nucleus, because the tail morphology clearly shows that radiation pressure is effective in accelerating the dust.  For example, even with zero ejection velocity, a constant radiation pressure acceleration of magnitude $\beta g_{\odot}$ (where $\beta$ is the dimensionless radiation pressure factor and $g_{\odot}$ is the gravitational acceleration to the Sun), would drive a grain to distance $\ell$ if

\begin{equation}
\beta = \frac{2 \ell R^2}{g_{\odot}(1) \tau^2}
\label{beta}
\end{equation}

\noindent  where $R$ is in AU, we have written $g_{\odot} = g_{\odot}(1) / R^2$ and $g_{\odot}(1)$ = 0.006 m s$^{-2}$ is the acceleration at $R = 1$ AU.  Parameter $\beta$ is approximately related to the particle size measured in microns, $a_{\mu m}$, by $\beta = a_{\mu m}^{-1}$.  Substituting $\ell = 7.3\times 10^5 \theta \Delta$, $\theta$ = 5.6\arcsec, $R$ = 2 AU and $\tau$ = 50 days into Equation (\ref{beta}) we obtain $\beta$ = 3$\times$10$^{-4}$, which corresponds roughly to a particle radius $a \sim$ 3.4 mm.  While this is clearly no better than an order of magnitude estimate, it is sufficient to show that the cross-section weighted size of the particles around the nucleus of 311P is large compared to the micron-sized dust which usually dominates the appearance of comets at this heliocentric distance.  

The mass of dust, $m_d$, and its total geometric cross-section, $C_e$, are related by

\begin{equation}
m_d = \frac{4}{3} \rho \overline{a} C_e
\end{equation}

\noindent where $\rho$ is the dust mass density and $\overline{a}$ is the effective mean radius of the dust particles, which we take to be $\overline{a} = 3.4$ mm based on Equation (\ref{beta}).  With $\rho$ = 3300 kg m$^{-3}$, we find a peak value $m_d$ = 3$\times$10$^6$ kg on UT 2013 September 23.  Expressed as a fraction of the nucleus mass, this is $f_d = m_d/M_n \sim$  3$\times$10$^{-5}$, showing that only a tiny fraction of the central mass has been shed.


\subsection{Plane-Crossing}

Earth crossed the orbital plane of 311P on UT 2013 December 31.08, permitting us to measure the extent of the dust perpendicular to the orbital plane, free from the effects of projection.  The resulting December 31 image shows the nucleus with a bright  tail consisting of dust ejected after UT 2013 August 21 extending to the east and a much fainter tail consisting of older dust to the west (Figure \ref{image}).  Since tails B -- G had largely dissipated already on UT 2013 December 08, we assume that most of the material seen to the east of the nucleus belongs to tail H, with a possible contribution from tail I.

\begin{figure}
\epsscale{1.0}
\begin{center}
\includegraphics[width=1.0\textwidth]{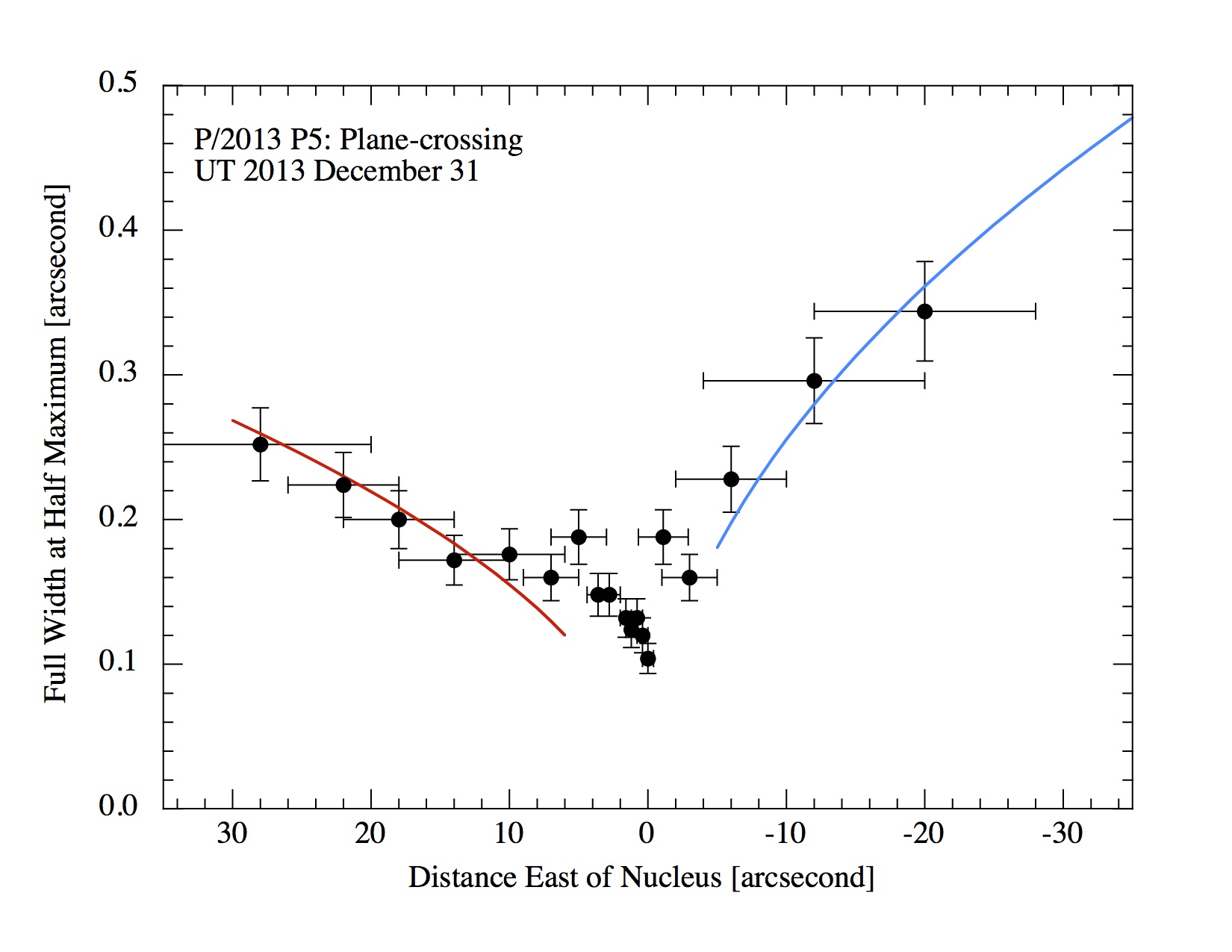}
\caption{Full width at half maximum of the 311P dust tail vs.~distance measured East from the nucleus from in-plane data taken UT 2013 December 31.  Vertical error bars show $\pm$10\% uncertainties on the FWHM measurements.  Horizontal bars show the range of distances along the dust tail over which each measurement was obtained.  Red and blue lines show $w \propto \ell^{1/2}$ relations fitted to the data.  \label{FWHM}} 
\end{center} 
\end{figure}

 Measurements of the Full Width at Half Maximum (FWHM) of the dust, $\theta_w$, were obtained by averaging over segments of the tail, increasing in size as distance from the nucleus increased and the surface brightness dropped (Figure \ref{FWHM}).  The measurements reveal an extraordinarily narrow tail, with $\theta_w <$ 0.3\arcsec~up to 30\arcsec~to the East of the nucleus and $\theta_w <$ 0.4\arcsec~up to 20\arcsec~to the West.  These widths correspond to physical FWHM $w_T$ = 436 km and 580 km, respectively.  

The motion of dust particles normal to the orbital plane is unaffected by radiation pressure, so that $w_T = 2 V_{\perp} t$, where $V_{\perp}$ is the perpendicular ejection velocity, $t$ is the time since ejection and the factor of two accounts for dust traveling both above and below the plane.  Radiation pressure acts in the orbital plane to push dust over the distance $\ell$ = $\beta g_{\odot}(1) t^2/(2 R^2)$, where $g_{\odot}$ is the gravitational acceleration to the Sun. Combining these expressions to eliminate $t$ we obtain

\begin{equation}
V_{\perp} =  \left[\frac{\beta g_{\odot}(1)}{8  \ell R^2}\right]^{1/2} w_T
\label{width}
\end{equation} 

\noindent Equation (\ref{width}) takes a particle size independent form when $V_{\perp} \propto a_{\mu m}^{-1/2}$, as is the case for particles ejected by the action of gas drag.  However, in 311P, gas drag is not likely to be responsible for the ejection of particles and so we cannot assume this inverse square-root relation.  To estimate a strong limit to $V_{\perp}$, we note that    $\beta \lesssim$ 1 for  wide range of even very small particles (Bohren and Huffman 1983).  Substituting $\beta \le$ 1 in Equation (\ref{width}), we obtain $v_{\perp} \le$ 0.8 m s$^{-1}$ to the east of the nucleus and $v_{\perp} \le$ 1.3 m s$^{-1}$ to the west.  Since the mean particle size in 311P is much larger than 1 $\mu$m (i.e.~$\beta \ll$ 1), we are justified to consider $\sim$1 m s$^{-1}$ to be a strong upper limit to $V_{\perp}$.   For example, taking $\overline{a}$ = 3.4 mm ($\beta$ = 3$\times$10$^{-4}$ as suggested in Section (\ref{quantity})), Equation (\ref{width}) gives $V_{\perp} \sim$ 3 cm s$^{-1}$.   Comparably small velocities, $\sim$3 to 7 cm s$^{-1}$, have been inferred by Moreno et al.~(2014).  Very small dust speeds are also indicated by the absence of a prominent coma in 311P (Figure \ref{image}).  

In Figure \ref{FWHM}, the FWHM of the tail projected to the East of the nucleus is well represented by $w \propto \ell^{1/2}$ for $\ell \gtrsim$ 12\arcsec~but the fit is less good for $\ell <$ 12\arcsec.  We suspect that the change in the trend of $w(\ell)$ is caused by overlap of tails H and I for $\ell <$ 12\arcsec.    In Section 4.5, we estimate the maximum length of tail I on UT 2013 December 31 as $\ell \sim$ 20\arcsec, supporting this interpretation.

\subsection{Dust Size Distribution}
The distance travelled by a dust grain of radius $a$ in a given time is proportional to the radiation pressure efficiency, $\beta$, while $\beta \propto 1/a$.  As a result, the surface brightness profile provides a measure of the dust size distribution relatively free from modeling uncertainties.

\begin{figure}
\epsscale{1.0}
\begin{center}
\includegraphics[width=1.0\textwidth ]{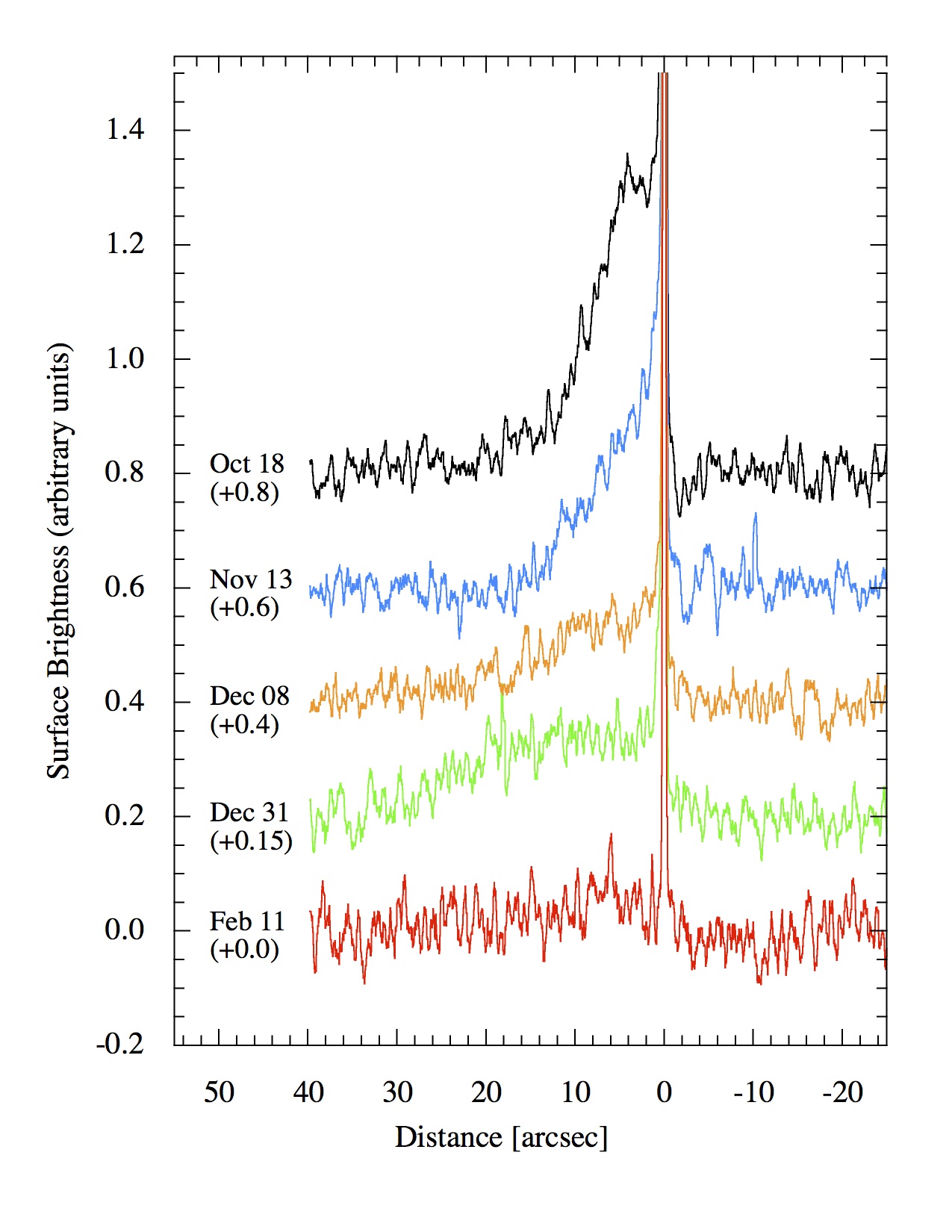}

\caption{Surface brightness (arbitrary scale) vs.~distance along the tail (positive distances are East of the nucleus).  The surface brightnesses have been smoothed along the tail direction by 0.44\arcsec~and shifted vertically by the indicated amounts for clarity of presentation.  \label{sb_profiles}} 
\end{center} 
\end{figure}

We measured the surface brightness profiles of the dust tails as a function of distance from the nucleus.   To determine tail surface brightness profiles, we used composite images constructed from the six images taken within a single orbit.  We first rotated each image about the nucleus position so as to bring the brightest tail to a horizontal orientation in the image plane.  Next, we used the average of the background pixels in boxes 1.0\arcsec~above and below the rotated tail to subtract residual gradients in the sky.  The sum of the counts within $\pm$0.5\arcsec~of the tail axis was then determined, initially retaining full (0.04\arcsec) resolution in the horizontal direction.  The extraction box is wide enough to extract the bulk of the tail signal without incurring large uncertainties due to the use of overly distant sky regions.  Extracted profiles were later smoothed along the tail axis direction where necessary, particularly in the later images in which the surface brightness was extremely low.  The extracted profiles (Figure \ref{sb_profiles}) show a characteristic asymmetric ramp up to the nucleus, but with evidence for changes between months that are larger than the scatter of the measurements.   The  surface brightness profiles in the Figure are shown smoothed by a 0.44\arcsec~running box average in order to decrease the noise in the data.  They are also offset vertically by the amounts indicated in the figure for clarity of presentation.  Angular distances are measured along the tail, increasing towards the East of the nucleus. Note that Figure (\ref{sb_profiles}) does not attempt to separate the contributions from different tails and is intended mainly to convey an impression of the decrease in the quantity of dust with time in 311P.

\begin{figure}
\epsscale{1.15}
\begin{center}
\includegraphics[width=1.15\textwidth ]{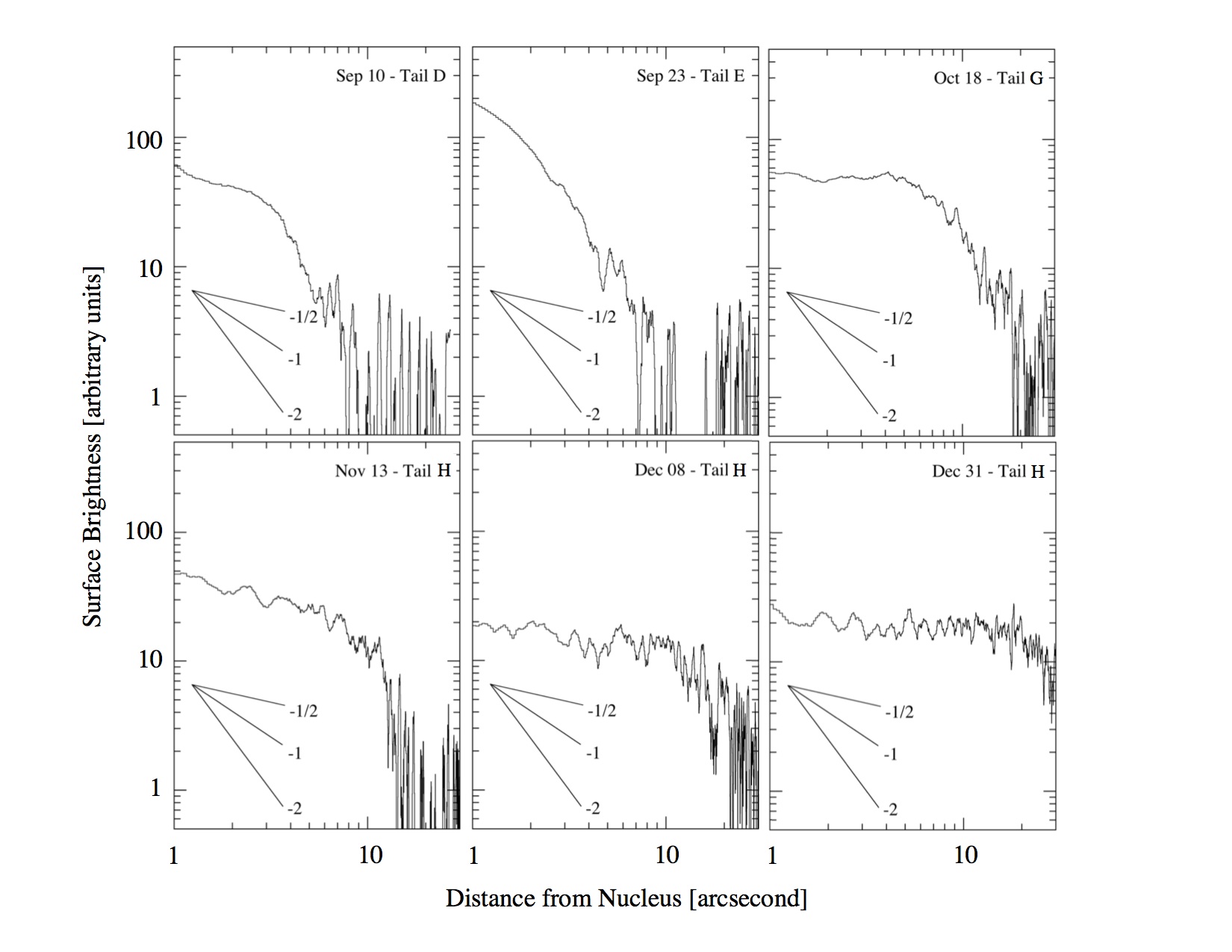}

\caption{Logarithmic plot of surface brightness (arbitrary scale) vs.~absolute angular distance along the tail.  The surface brightnesses have been smoothed along the tail direction by 0.44\arcsec.  Power law slopes of -1/2, -1 and -2 are shown for reference.  \label{tail_profiles}} 
\end{center} 
\end{figure}

Indeed, even at the resolution offered by HST, most tails are overlapped by other tails or background features, compromising the extraction of individual profiles.  However, Tail D on September 10 and Tail E on September 23 could be photometrically isolated from other tails with a reasonable degree of confidence, and we were able to measure Tail H  on three occasions in November and December (Figure \ref{tail_profiles}).   

Figure (\ref{tail_profiles}) shows the measured surface brightness profiles together on a common, logarithmic scale.  We have excluded the region within 1\arcsec~of the nucleus to avoid complications caused by the central nucleus, and regions beyond 30\arcsec~are not plotted because of inadequate signal-to-noise ratios.   The surface brightness profiles are clearly not all power law functions of distance from the nucleus.  Nevertheless, it is convenient to discuss the profiles in terms of power law relationships, in which the surface brightness, $\Sigma(\ell)$, at distance $\ell$ from the nucleus is $\Sigma(\ell) \propto \ell^{-s}$, where $s$ is a constant.  Lines indicating gradients $s$ = 1/2, 1 and 2 have been added to Figure (\ref{tail_profiles}) to guide the eye.  

The tail profiles show a range of shapes. Evidently, there are considerable differences between the values of $s$ pertaining to different tails, ranging from steep $s \sim$ 2.1 (Tail E on September 23) to shallow $s \sim$ 0.1 (Tail H on December 31).  When plotted as a function of the tail age (using the best-fit solutions summarized in Table \ref{angles}), it appears that $s$ shows a trend towards smaller values with increasing age.  This is most noticeable in Tail H, for which $s$ = 0.6, 0.3 and 0.1 on November 13, December 08 and December 31, respectively, but it is also true of the younger tails D, E and G shown in Figure (\ref{tail_profiles}). The profile on December 31 is likely an overlap of tails H and I, which may explain its particularly low value of $s$.

\begin{figure}
\epsscale{0.99}
\begin{center}
\plotone{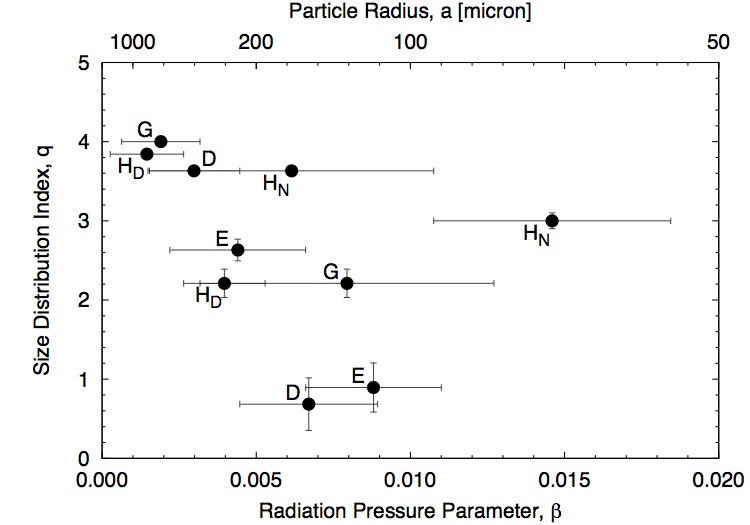}
\caption{Size distribution exponent $q$ as a function of radiation pressure coefficient $\beta$ determined from fits to the data in Figure (\ref{tail_profiles}). The horizontal error bars  indicate the range of  $\beta$ value of the particles used in the fit, while the vertical error bars indicate the uncertainty in $q$. The upper horizontal axis shows the approximate particle radius computed from $a = 1/\beta$.  Labels  refer to particular tails as identified in Figure \ref{image_arrows}, with subscripts N and D to distinguish separate measurements of tail H taken in November and December, respectively. \label{slope}} 
\end{center} 
\end{figure}

If the radii of ejected particles, $a$, are distributed according to a power-law relation $n(a)da \propto a^{-q} da$, where $n(a)da$ is the number of particles having radii in the range $a$ to $a + da$, then the size index $q$ and the surface brightness index $s$ are related, in the geometric optics limit, by $q$ = 4 - $s$.  The flatter surface brightness profiles in Figure (\ref{tail_profiles}) indicate larger relative proportions of small particles.  For example, the three profiles of Tail H in the Figure correspond to a size index that steepens from $q$ = 3.4 (November 13), to 3.7 (December 08) to 3.9 (December 31).   However, even as the proportion of small particles increases, the mean size of the particles at a given distance from the nucleus grows with the square of the time since ejection. For example, on the above-mentioned three dates, the approximate size of the dust measured 10\arcsec~from the nucleus increases from 65 $\mu$m to 260 $\mu$m to 380 $\mu$m (values from synchrone of H) as a result of outward sweeping by radiation pressure.  Therefore, the surface profiles in Figure (\ref{tail_profiles}) indicate that the size distribution itself steepens toward larger particle sizes; $s$ is  a function of $a$. Figure~(\ref{slope}) shows the dependence of $q$ on $\beta$, measured from the surface brightness profiles in Fig.~\ref{tail_profiles}, where $\beta$ is proportional to the nucleus distance for a given tail. The scatter in  Figure~(\ref{slope}) reflects the wide range of surface brightness gradients shown by different tails, but the Figure suggests a trend such that the size distribution is steeper (larger $q$)  for smaller $\beta$ (larger particles).

It is interesting to note that in-situ observations of the S-type  rubble pile asteroid Itokawa show a qualitatively similar steepening of the size distribution with increasing particle size.  Specifically, regolith particles with diameters from 30 $\mu$m to 180 $\mu$m follow a differential power law size distribution $q = 3.0$ (Tsuchiyama et al.~2014), while large blocks with diameters $D \ge$ 5 m  follow $q = 4.1\pm0.1$ (Michikami et al.~2008).   Small particles on Itokawa are presumably created by micrometeorite bombardment of the exposed surface. The large blocks are too numerous to have been formed by the few known impact craters on this body and are instead thought to be products of a past shattering collision followed by re-accretion under self-gravity.  Perhaps a similar explanation is responsible for the size dependence of $s$ implied by the surface brightness profiles of the 311P dust tails.

\subsection{Dust size range}
\label{sizes}
\begin{figure}
\epsscale{0.99}
\begin{center}
\includegraphics[width=0.8\textwidth, angle =270 ]{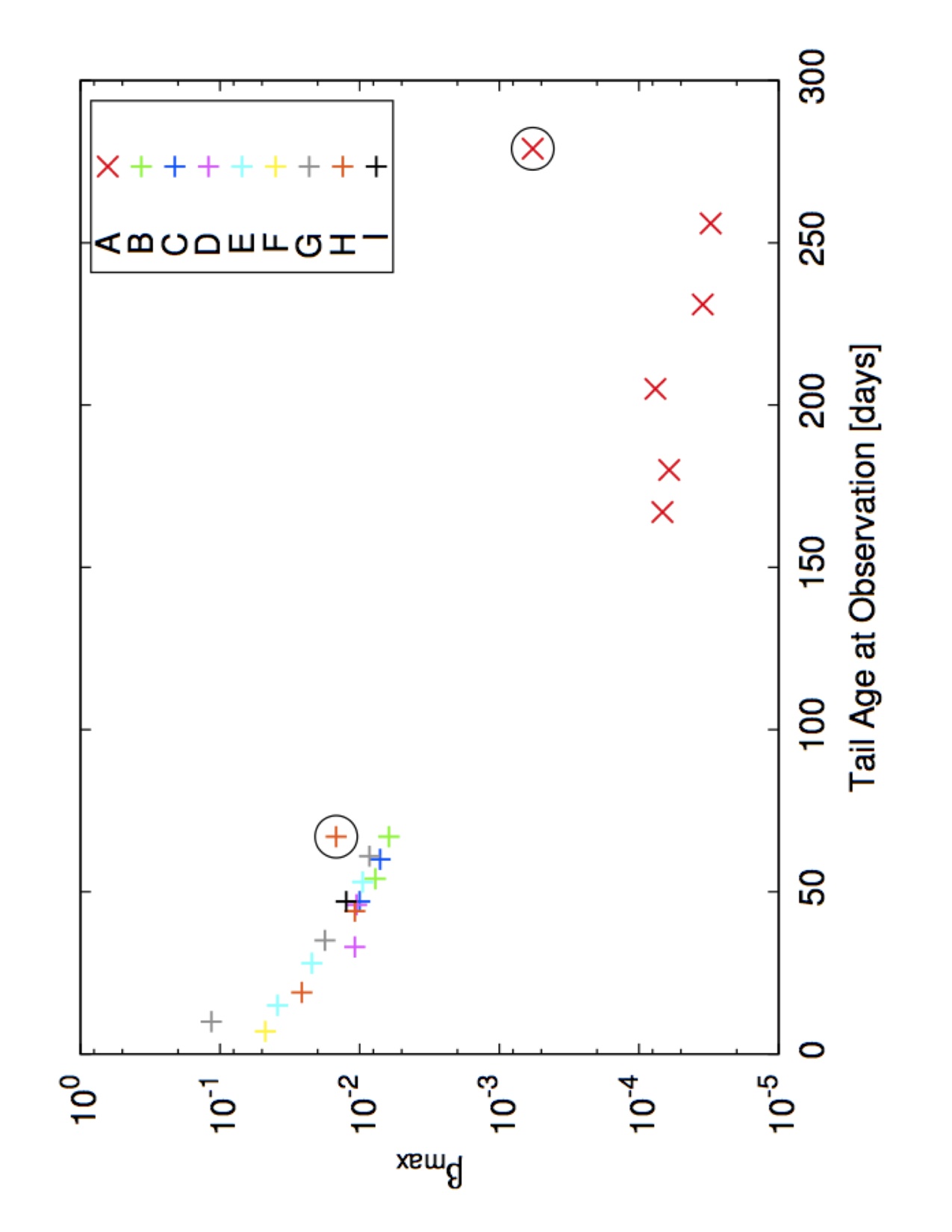}

\caption{Maximum radiation pressure factor, $\beta_{max}$, inferred from the tail length and plotted as a function of the age of the tail. Measurements from December 31, thought to be affected by overlap of tails as the Earth crossed the orbital plane, are circled.  \label{betamax}} 
\end{center} 
\end{figure}

In a synchronic tail (i.e.~with all particles ejected at the same time) the length of the tail gives a measure of the largest $\beta$ (corresponding to the smallest $a$, because $\beta \propto a^{-1}$) in the distribution of ejected particles.  We refer to this largest $\beta$ as $\beta_{\max}$. We derived $\beta_{\max}$ for each tail at each epoch of observation and show the result in  Figure~\ref{betamax} as a function of the instantaneous age of the tail. Evidently, $\beta_{\max}$ decreases with increasing tail age, because larger and larger particles are pushed away from the nucleus by radiation pressure, decreasing the surface brightness and signal-to-noise ratio.  The relation between $\beta_{\max}$ and age is very similar for all tails (Figure \ref{betamax}), indicating that the total amount of dust in each tail is also similar. The main exceptions to this are the two tails observed during the plane crossing on December 31 (marked in the Figure), which are both detected over significantly longer distances than expected from the other dates of observation. The likely reason is that projection effects, due to overlapping tails in the orbital plane of the object, artificially extend the distance to which dust can be detected. 

The largest values of $\beta_{\max} \sim 0.1$ (smallest particles) are found in the youngest tails, corresponding to particle radii $a \sim$ 10 $\mu$m.  The apparent absence of particles smaller than 10 $\mu$m is striking, given that the scattering cross-section in comets is normally dominated by particles with a radius comparable to the wavelength of observation.  The depletion of the smallest particles could be a result of grain clumping by van der Waals forces in the material sloughed off from the nucleus (Marshall et al.~2011).  Assuming that the maximum detectable $\beta$ was 0.1 also for tail I, we infer from the error bars on the ejection date (December 26 $\pm$ 5 days) that during the plane crossing observation, the length of I may have been up to 20~arcsec, which is consistent with the break in the FWHM-profile at a nucleus distance of 12~arcsec seen in Figure~\ref{FWHM} and discussed in Section 4.3.

Figure~\ref{betamax} also shows that tail A is distinguished by consisting of old and large particles, having  $\beta_{\max} \sim 3 \times 10^{-5}$ to 7 $\times 10^{-5}$ (radii $\sim$15 to 30~mm), with even larger particles located closer to the nucleus. The non-detection of a gap between tail A and the nucleus on December 08 indicates that dust was present at least up to 1 arcsec from the nucleus. The corresponding $\beta = 1.2 \times 10^{-5}$ means that particles as large as radius $a \sim$  80~mm must be present in the ejected dust.

\section{Discussion}

The results derived here from the full HST data set generally support and strengthen the conclusions reached in Paper 1 from only the first two HST orbits.  The drawn-out series of impulsive mass loss events recorded by the individual tails of 311P rule out impact as a cause.  Impact, as observed in (596) Scheila (Bodewits et al.~2011, Jewitt et al.~2011, Ishiguro et al.~2011) produces a single burst of ejecta which dissipates monotonically by radiation pressure sweeping.  It provides no mechanism for the repeated ejection of dust in bursts spread over many months.  Sublimation of ice constitutes an equally unconvincing explanation of the observations. Comets that are known to contain sublimating ice release dust continuously,  not  in intermittent bursts like 311P.  Furthermore, 311P resembles members of the Flora asteroid family, both in its orbit at the inner edge of the asteroid belt and in its optical colors, which are those of an S-type asteroid (Paper 1, Hainaut et al.~2014).  The mineralogy of the Floras  is thought to reflect high temperature metamorphic processes that are incompatible with the survival of water ice.

Unlike the clearly fragmented  P/2013 R3 (Jewitt et al.~2014), the body of 311P lacks discrete companions and shows no evidence of having disrupted as a result of its own spin.  Indeed, the small value of the ejecta to nucleus mass ratio, $f_d \sim 3\times10^{-5}$, shows that a negligible fraction of the central body of 311P has been lost, unlike in P/2013 R3 where multiple 200 m scale bodies are observed slowly separating.  Nevertheless, we infer that rotation is plausibly implicated in the mass loss from 311P, for three reasons.  First, the central light photometry (Figure \ref{DV_Dt}) shows variations on sub-hour timescales that are compatible with rapid nucleus rotation.  Second, the low debris ejection speeds inferred from the orbital plane crossing observations are exactly as expected in rotationally induced mass loss (in which the ejection speed and the gravitational escape speed should be of the same order, as measured).  Third, the small size of the nucleus (estimated nucleus radius $r_n \le 200$ m), suggests the YORP torque as a plausible agent by which to accelerate the spin (Marzari et al.~2011).  While none of these facts amounts to proof that 311P is losing mass rotationally, they are individually consistent with this interpretation and collectively suggestive of it.  We note that, in addition to the clear-cut case of P/2013 R3, rotation has also been implicated in the case of active asteroid P/2010 A2 (Agarwal et al.~2013) and in 133P/Elst-Pizarro (Jewitt et al.~2014b).  In the latter case, mass loss driven by sublimation at four consecutive perihelia appears to be assisted by centripetal acceleration owing to  rotation of the elongated nucleus with a $\sim$3.6 hr period.

Hirabayashi and Scheeres (2014) defined ``mass shedding'' as local loss of weakly bonded regolith without corresponding structural instability of the underlying asteroid.  Representing the nucleus as a sphere of radius $r_n$, with fractional mass loss $f_d \sim 3\times10^{-5}$, we find that the ejected material corresponds to a global surface layer of thickness $\Delta r = f_d (r_n/3)$, or $\Delta r \sim$ 2 mm.   More likely, mass is shed from thicker regions occupying  a smaller fraction of the surface, where materials are locally unstable to equator-ward movement and detachment from the asteroid.   Once detached, avalanched debris is picked up by solar radiation pressure to form the observed tails.  Repeated small shedding events, together with continued action of YORP torques, might hold the central body close to rotational instability for a long time.  If we assume that 311P lost a fraction of the nucleus mass, $f_d$, in the past $\sim$1 yr, the time taken to lose all the mass would be $\tau \sim f_d^{-1} = 3\times10^4$ yr. 

This appealing picture raises many questions.  For example, does each episode of mass loss on 311P correspond to a separate avalanche of weakly bonded debris from the surface, or are successive avalanches related, with one triggering another?    What determines the mass of material lost in an avalanche?  Does the mass loss follow, for example, the type of power-law relation observed in the collapse of self-organized critical sand-piles (Laurson et al.~2005)? What sets the duration between mass loss events (i.e.~why weeks and months (Table \ref{angles}), not hours or thousands of years)?  For how long will these events persist?  Will continued mass loss  eventually lead to the formation of a close binary (c.f.~Jacobsen et al.,~2011; Walsh et al. 2012)? Or has it already done so, with the secondary orbiting in the unresolved core of the HST image?  Will the main mass eventually  be driven to structural instability as in P/2013 R3 (Jewitt et al.~2014; Hirabayashi et al.~2014)? Or could 311P be a relic of structural instability occurring in a long-ago disrupted precursor body?   While it might not be possible to answer all these questions,  we are hopeful that new observations,  coupled with improved models, will  lead to a better understanding of  this most exciting object and of asteroid disruption generally.


\clearpage

\section{Summary}

We used the Hubble Space Telescope to study active asteroid 311P/PANSTARRS (2013 P5) with the following results.  

\begin{enumerate}

\item The faintest measured absolute magnitude of the nucleus, $H_V$ = 18.98$\pm$0.10 (S-type phase function assumed), corresponds to a mean radius of 0.20$\pm$0.02 km at the  assumed geometric albedo $p_V$ = 0.29$\pm$0.09.  No companion nuclei brighter than $V$ = 28.5 are detected beyond 0.4\arcsec~(370 km) from the nucleus. The radius of any such companion must be $<$10 m, assuming the same albedo and phase function.

\item The nucleus shows brightness variations up to $\sim$0.1 to 0.15 magnitudes that occur on timescales shorter than the aperture crossing times for ejected dust.    These variations are likely related to the rotation of the nucleus. 

\item The nucleus is an intermittent source of dust which, once released, is pushed by solar radiation pressure into distinct tails defined by their projected position angle on the sky and, through a model, by their emission date.  The first (UT 2013 March 27) and last (UT 2013 December 26) ejected dust tails in our data differ in age by 274 days ($\sim$9 months).

 \item Slow fading (timescale $\gtrsim$50 days) of the integrated dust component of 311P argues for a very low mean dust velocity, $v \lesssim$ 1 m s$^{-1}$.  If attributed to the action of radiation pressure on particles ejected with negligible initial speed, this implies a mean radius $\sim$3 mm and a total ejected dust mass $\sim 3\times$10$^6$ kg, or about 3$\times$10$^{-5}$ of the mass of the central nucleus. 
  
\item The narrow extent of the dust tail perpendicular to the orbital plane further implies very low ejection speeds, $\sim$0.06 m s$^{-1}$ for 3 millimeter radius particles.  In comparison, the gravitational escape speed of a non-rotating, 200 m radius spherical body of density 3300 kg m$^{-3}$ is  0.27 m s$^{-1}$.

\item  The episodic ejection of debris at sub-meter per second velocities, the detection of short term ($<$0.7 hr) nucleus brightness variations and the absence of discrete secondaries are all consistent with mass shedding from an asteroid driven to rotational instability.

\end{enumerate}

\acknowledgments
We thank Masateru Ishiguro, Toshi Hirabayashi, Man-To Hui and the anonymous referee for comments.  Based on observations made with the NASA/ESA \emph{Hubble Space Telescope,} with data obtained at the  Space Telescope Science Institute (STScI).  Support for programs 13475 and 13609 was provided by NASA through a grant from the Space Telescope Science Institute, which is operated by the Association of Universities for Research in Astronomy, Inc., under NASA contract NAS 5-26555.  We thank Alison Vick, Tomas Dahlen, and other members of the STScI ground system team for their expert help in planning and scheduling these Target of Opportunity observations.

\clearpage

\clearpage

\begin{deluxetable}{lllllllr}
\tablecaption{Observing Geometry 
\label{geometry}}
\tablewidth{0pt}
\tablehead{
\colhead{UT Date and Time}  & DOY\tablenotemark{a} & \colhead{$R$\tablenotemark{b}}  & \colhead{$\Delta$\tablenotemark{c}} & \colhead{$\alpha$\tablenotemark{d}}   & \colhead{$\theta_{\odot}$\tablenotemark{e}} &   \colhead{$\theta_{-v}$\tablenotemark{f}}  & \colhead{$\delta_{\oplus}$\tablenotemark{g}}   }
\startdata
2013 Sep  10 16:44 - 17:24 & 253& 2.112 & 1.115  & 5.1 & 125.0 & 244.8  & -4.2\\
2013 Sep  23 09:20 - 09:59 & 266 & 2.096 & 1.135  & 10.7 & 89.2 & 244.5  & -4.3\\
2013 Oct 18    15:15 - 15:55  & 291 &  2.064 & 1.278  & 21.7  &  75.4 & 244.1    & -3.72\\
2013 Nov 13  06:39 - 07:20  & 317 &   2.033           &    1.505         &    27.7  &     70.39       &    244.2     &   -2.44    \\
2013 Dec 08  07:46 - 08:27  & 342 &   2.006  &  1.758  &  29.4   &  67.28  &   244.6     &  -1.09  \\
2013 Dec 31  02:35 - 03:15   & 365 &  1.985   &  1.991   & 28.6  &  65.39   &  245.4     &  +0.00  \\
2014 Feb 11   09:59 - 10:40  & 407 &    1.954          &    2.381         &    23.8      &     64.47       &    248.4           &   +1.40    \\

\enddata

\tablenotetext{a}{Day of Year, 2013}
\tablenotetext{b}{Heliocentric distance, in AU}
\tablenotetext{c}{Geocentric distance, in AU}
\tablenotetext{d}{Phase angle, in degrees}
\tablenotetext{e}{Position angle of the projected anti-Solar direction, in degrees}
\tablenotetext{f}{Position angle of the projected negative heliocentric velocity vector, in degrees}
\tablenotetext{g}{Angle of Earth above the orbital plane, in degrees}

\end{deluxetable}

\clearpage

\begin{deluxetable}{lccccc}
\tablecaption{Nucleus Photometry 
\label{photometry}}
\tablewidth{0pt}
\tablehead{
\colhead{Date}      & DOY\tablenotemark{a} &  \colhead{$V_{0.2}$\tablenotemark{b}} & \colhead{$H_{0.2}(C)$\tablenotemark{c}} & \colhead{$H_{0.2}(S)$\tablenotemark{d}} }
\startdata
2013 Sep 10     & 253  & 20.92$\pm$0.01  	& 18.63 & 18.69  \\
2013 Sep 23     & 266 & 21.01$\pm$0.01 		& 18.46 & 18.54   \\
2013 Oct 18      & 291 &  21.95$\pm$0.01  	& 18.79 &   18.91         \\    
2013 Nov 13    &  317 &  22.36$\pm$0.01 	& 18.70 & 18.84            \\   
2013 Dec 08    & 342  &  22.68$\pm$0.01 	&  18.66 & 18.81           \\ 
2013 Dec 31    & 365 & 22.84$\pm$0.01   	&  18.60 &  18.75            \\     
2014 Feb 11   & 407  &      23.30$\pm$0.02 	& 18.84 & 18.98                \\   

\enddata

\tablenotetext{a}{Day of Year, 2013}
\tablenotetext{b}{Apparent V magnitude within 5 pixel (0.2\arcsec) radius aperture.  Quoted uncertainty is the statistical error, only.}
\tablenotetext{c}{Absolute V magnitude computed from $V_{0.2}$ assuming a C-type phase function.}
\tablenotetext{d}{Absolute V magnitude computed from $V_{0.2}$ assuming an S-type phase function.}

\end{deluxetable}

\clearpage

\begin{deluxetable}{lccccrr}
\tablecaption{Dust Photometry 
\label{total_photometry}}
\tablewidth{0pt}
\tablehead{
\colhead{Date}      & DOY\tablenotemark{a}  &  \colhead{$\theta$\tablenotemark{b}}&  \colhead{$V_{\theta}$\tablenotemark{c}} & \colhead{$H_{\theta}(S)$\tablenotemark{d}} & \colhead{$H_{dust}$\tablenotemark{e}}  & \colhead{$C_e$\tablenotemark{f}} }
\startdata
2013 Sep 10     & 253  &  	6.0	& 20.19$\pm$0.03  	& 17.96 	&  18.74	& 0.16  \\
2013 Sep 23     & 266 &  	5.9	& 20.21$\pm$0.03 		& 17.74  	& 18.44	& 0.20   \\
2013 Oct 18      & 291 &  	5.2	&  21.18$\pm$0.03  	& 18.14  	& 	18.88 &  0.14         \\    
2013 Nov 13    &  317 &  	4.4	&  21.85$\pm$0.05 	& 18.33  	& 19.40	& 0.08            \\   
2013 Dec 08    & 342  &  	3.8	&  22.10$\pm$0.05 	&  18.23  	& 	19.20 & 0.10          \\ 
2013 Dec 31    & 365 &  	3.4	& 22.23$\pm$0.05   	&  18.14  	& 19.05 	& 0.12            \\     
2014 Feb 11   & 407  &  	2.8	&      23.33$\pm$0.10	& 19.01  	& $>$19.5 	& $<$0.07               \\   

\enddata

\tablenotetext{a}{Day of Year, 2013}
\tablenotetext{b}{Angular radius of photometry aperture, in arcsecond}
\tablenotetext{c}{Apparent V magnitude within  aperture radius $\theta$\arcsec.  Quoted uncertainty is the statistical error, only.}
\tablenotetext{d}{Absolute V magnitude within  aperture radius $\theta$\arcsec.  }
\tablenotetext{e}{Absolute V magnitude of dust alone, computed from Equation (\ref{dust})}
\tablenotetext{f}{Effective cross-section of the dust, in km$^2$, from Equation (\ref{invsq})}

\end{deluxetable}

\clearpage

\begin{landscape}
\begin{deluxetable}{lcccccccccc}
\tablecaption{Tail Position Angles 
\label{angles}}
\tablewidth{0pt}
\tablehead{\colhead{UT Date\tablenotemark{a}}  &
\colhead{DOY\tablenotemark{b}}  & \colhead{A}    & \colhead{B}   & \colhead{C} & \colhead{D} & \colhead{E} & \colhead{F} & \colhead{G} & \colhead{H} & \colhead{I}}
\startdata

2013 Sep 10 & 253 	& 237$\pm$1 		& 220$\pm$1 & 216$\pm$2  & 202$\pm$1   & 161$\pm$2 & 141$\pm$2 & -- & -- & -- \\
2013 Sep 23 & 266 	& 234$\pm$1 		& 198$\pm$1 & 190$\pm$2  & 153$\pm$2  & 114$\pm$1 & -- & 97$\pm$2 & -- & --   \\
2013 Oct 18 & 291 	& 234$\pm$2 		&  -- &  -- &--  & 86.5$\pm$0.5  & -- & 80.5$\pm$0.5 & -- & --\\
2013 Nov 13 & 317 	&  235.5$\pm$2.5	&  -- &  --  & -- & -- & -- & 74.8$\pm$0.1 &  71.9$\pm$0.1 & -- \\
2013 Dec 08 & 342 	&  244.1$\pm$1.0 	& -- &  -- & -- & --& -- &-- & 68.3$\pm$0.1& -- \\
2013 Dec 31 & 365 	& 245.5$\pm$0.1 	&  -- &   & -- & -- & --  & -- & 65.5$\pm$0.1  & --\\
2014 Feb 11 & 407 	& -- &  -- & -- & -- & --  & -- &  -- & -- & 61.1$\pm$0.6\\
\hline
{Best Fit Date}\tablenotemark{c} &    &   Mar 27  &  Jul 18   &   Jul 25 &  Aug 08 &  Aug 26 &  Sep 03 &  Sep 13  & Oct 25  & Dec 26\\
{DOY}\tablenotemark{d}               &    &   86  &  199   &   204 &  220 &  238 &  246 &  256  & 298  & 360\\

\enddata

\tablenotetext{a}{Calender date of observation}
\tablenotetext{b}{Corresponding Day of Year, 2013}
\tablenotetext{c}{Date of the synchrone best matching the measured position angles}
\tablenotetext{d}{Day of Year of the synchrone best matching the measured position angles}

\end{deluxetable}
\end{landscape}

\clearpage

\clearpage

%
%

\clearpage

\clearpage

\clearpage

\clearpage

\clearpage

\clearpage

\clearpage

\clearpage

\clearpage

\clearpage

\clearpage

%
%
%

\end{document}